\documentstyle[aps,prl,epsf,floats]{revtex}
\newcommand{\beq}{\begin{equation}}
\newcommand{\eeq}{\end{equation}}
\newcommand{\beqs}{\begin{eqnarray}}
\newcommand{\eeqs}{\end{eqnarray}}

\newcommand{\lsim}{\mathrel{\raisebox{-.6ex}{$\stackrel{\textstyle<}{\sim}$}}}
\newcommand{\gsim}{\mathrel{\raisebox{-.6ex}{$\stackrel{\textstyle>}{\sim}$}}}

\draft

\tighten

\begin{document}

\twocolumn[\hsize\textwidth\columnwidth\hsize\csname
@twocolumnfalse\endcsname

\title{Lepton Dipole Moments in Extended Technicolor Models}

\author{Thomas Appelquist$^1$, Maurizio Piai$^1$, and Robert Shrock$^2$}
\address{$^1$ Department of Physics, Sloane Laboratory, Yale University,
New Haven, CT 06520 \\ C. N. Yang Institute for Theoretical
Physics, State University of New York, Stony Brook, NY  11794}

\baselineskip 6.0mm

\tighten

\maketitle

\begin{abstract}

We analyze the diagonal and transition magnetic and electric dipole moments
of charged leptons in extended technicolor (ETC) models, taking account of
the multiscale nature of the ETC gauge symmetry breaking, conformal
(walking) behavior of the technicolor theory, and mixing in the
charged-lepton mass matrix. We show that mixing effects dominate the ETC
contributions to charged lepton electric dipole moments and that these can
yield a value of $|d_e|$ comparable to the current limit. The rate for $\mu
\to e\gamma$ can also be close to its limit.  From these and other processes
we derive constraints on the charged lepton mixing angles. The constraints
are such that the ETC contribution to the muon anomalous magnetic moment,
which includes a significant lepton mixing term, can approach, but does not
exceed, the current sensitivity level.

\end{abstract}

\pacs{12.15.Ff, 12.60.Nz, 13.40.Em, 13.40.Hq}

\vskip2.0pc]

We study the magnetic and electric dipole moments of charged leptons in a class
of extended technicolor (ETC) models \cite{at94}-\cite{ckm}.  We also analyze
the transition moments and the resultant electromagnetic decays.  Charged
lepton mixing plays a crucial role in determining each electric dipole moment
(EDM). For the electron, the EDM can be comparable to the current experimental
upper limit. Bounds on charged lepton mixing are derived from the constraint
that the electron EDM be smaller than this limit and from upper limits on the
decays $\mu \to e \gamma$, $\tau \to (e,\mu)\gamma$.

In technicolor theories, electroweak symmetry breaking (EWSB) arises from a
new, strongly coupled gauge interaction at TeV energy scales \cite{tc}.  Quark
and lepton mass matrices arise from the embedding of technicolor in a larger
gauge theory, extended technicolor (ETC)\cite{etc}, which must break
sequentially as the energy decreases from energies on the order of $10^3$ TeV
down to the TeV level. Precision measurements place tight constraints on these
theories, suggesting a small number of new degrees of freedom at the TeV scale
and non-QCD-like behavior of the technicolor theory. With this motivation, some
attention has been focused on walking technicolor theories, which exhibit an
approximate conformal behavior, with large anomalous dimensions, in the
infrared \cite{wtc}.

While this provides an attractive framework, it has been a challenge to
construct explicit models along these lines, with, for example, the necessary
ingredients to effect the requisite ETC symmetry breaking at each stage. In
Refs. \cite{at94,nt,lrs,ckm} a class of ETC models was developed which has
these ingredients.  The models are based on an ETC gauge group SU(5) which
commutes with the standard-model gauge group and breaks in stages
corresponding to the three fermion generations, to a residual SU(2)$_{TC}$
technicolor gauge theory, naturally producing a hierarchy of charged lepton and
quark masses. The models also exhibit charged-current flavor mixing,
intra-family mass splittings, a dynamical origin of CP-violating phases in the
quark and lepton sectors, and a see-saw mechanism for light neutrinos without
the presence of a grand unified scale \cite{nt}. The choice SU(2)$_{TC}$ (i)
minimizes the TC contributions to the electroweak $S$ parameter, (ii) with a
standard-model family of technifermions in the fundamental representation of
SU(2)$_{TC}$, can yield an approximate infrared fixed point and associated
walking behavior, and (iii) makes possible the mechanism of \cite{nt}
explaining light neutrinos.

The breaking to SU(2)$_{TC}$ is driven by the ETC interaction itself, a
chiral gauge theory, along with one additional, strong SU(2) gauge
interaction. Each requisite breaking is shown to be plausible, within
strong-interaction uncertainties. A set of standard-model-singlet fermions,
including right-handed neutrinos, is naturally part of the models, and it is
these particles that condense at the various ETC breaking scales, producing
the ETC gauge boson masses and mixing, and leaving the residual unbroken
SU(2)$_{TC}$ technicolor theory.  At the scale $\Lambda_{TC}$, technifermion
condensates break the electroweak symmetry, yielding $m_W^2 = (g^2/4) f_F^2
(N_c+1)$ where $g$ is the SU(2)$_L$ gauge coupling, $f_F \simeq 130$ GeV,
and $\Lambda_{TC} \simeq 300$ GeV.

The class \cite{at94,nt,lrs,ckm} has not yet led to a fully realistic model,
yielding, for example, the measured fermion masses and mixings. Also, the
models have a small number of phenomenologically unacceptable Nambu-Goldstone
bosons, arising from spontaneously broken U(1) global symmetries \cite{ckm}. To
give them masses above current bounds, some additional interactions that
explicitly break the U(1) symmetries must be invoked at energies above the
highest ETC scales. However, all the models share interesting generic
features, including a mechanism for CP violation, that are worth studying in
their own right.

The bilinear fermion condensates at each stage of ETC breaking have, in
general, non-vanishing phases, providing a natural, dynamical source of CP
violation. The underlying theory consists of massless fermions and gauge
fields, and is free of gauge anomalies. There are global chiral symmetries,
some of which are anomalous, and hence are broken by instantons. The $F_{\mu
\nu} \tilde{F}^{\mu\nu}$ terms associated with each (nonabelian) gauge
interaction may be rotated away by chiral transformations through the relevant
global anomalies. The phases that develop at each stage of ETC breaking should
be calculable \cite{lane_cpv} in a non-perturbative treatment of a fully
realistic model. Here we take the phases to be arbitrary. Phases of order unity
seem natural in our context and are consistent with the fact that in the quark
sector the CP-violating quantity $ \sin \delta$ in the CKM matrix is not small.

Below the electroweak breaking scale, the effective theory includes the
standard-model interactions, dimension-3 mass terms for the quarks and charged
leptons, a dimension-3 mass term for the electroweak-doublet and
standard-model-singlet neutrinos, and a tower of operators of dimension-5 and
higher describing the new physics of the model(s). The mass matrices are
complex and have both diagonal and off-diagonal entries, inheriting these
features from the ETC gauge boson mixings and phases arising at each breaking
stage. Among the dimension-5 operators is one describing the electric and
magnetic dipole moments of the charged leptons. It, too, involves a complex
matrix with diagonal and off-diagonal entries. The QCD interactions include a
$G_{\mu \nu} \tilde{G}^{\mu\nu}$ term with its associated strong CP problem, as
well as a dimension-5 term describing chromo-magnetic and chromo-electric
dipole operators for the quarks. Whether a resolution of the strong CP problem
will emerge in the class of models \cite{at94,nt,lrs,ckm} is not yet clear
\cite{lane_cpv}. In this letter, we focus on the charged lepton sector,
including its CP violation.

The charged lepton mass matrix $M^{(\ell)}$ appears in the dimension-3
operator
\beq
{\cal L}_m = -\bar{\ell}_{L,j} M^{(\ell)}_{jk} \ell_{R,k} + h.c.
\label{leptonmass}
\eeq
where $\ell=(e,\mu,\tau)$ denotes the technisinglet ETC interaction
eigenstates. We will estimate $M^{(\ell)}$ from the underlying ETC theory.
It can be brought to real, diagonal form by the bi-unitary transformation
\cite{msym}
\beq
U_L M^{(\ell)} U_R^{-1} = M^{(\ell)}_d \ .
\label{biunitary}
\eeq
Hence, the interaction eigenstates are mapped to mass eigenstates $\psi$ via
$\ell_L = U_L^{-1} \psi_L$, $\ell_R = U_R^{-1} \psi_R$.

The magnetic and electric dipole matrices of the charged leptons
are given by the dimension-5 operator
\beq
{\cal L}_{dip.}=\frac{1}{2} \bar \ell_{j,L} \tilde D_{jk} \sigma_{\mu\nu}
\ell_{k,R} F^{\mu\nu}_{em} + h.c. \label{dipoleop} \eeq
The matrix $\tilde D$ will also be estimated from the underlying ETC theory.
In terms of mass eigenstates,
\beq
\bar \ell_L \tilde D \sigma_{\mu\nu} \ell_R F^{\mu\nu}_{em} +
h.c. = \bar \psi_L D \sigma_{\mu\nu} \psi_R F^{\mu\nu}_{em} + h.c.,
\label{dipoleopuu} \eeq
where
\beq
D = U_L \tilde D U_R^{-1}  \ .
\label{ddtilderel} \eeq
Decomposing $D$ into hermitian and anti-hermitian parts, $D = D_H
+ D_{AH}$, where $D_H = (D+D^\dagger)/2$ and $D_{AH} =
(D-D^\dagger)/2$, the dipole operator is
\beq
\frac{1}{2} \Bigl [ \bar \psi D_H \sigma_{\mu\nu} \psi + \bar
\psi D_{AH} \sigma_{\mu\nu} \gamma_5 \psi \Bigr ] F^{\mu\nu}_{em} \ .
\label{dhdah} \eeq
Then $a_{\psi_j} = (g_{\psi_j}-2)/2 = (2m_{\psi_j}/e)D_{H,jj}$ (where
$e=-|e|$ is the lepton charge) and $d_{\psi_j} = -iD_{AH,jj} $ \cite{ls}.

The $M$ and $D$ matrices, arising from physics at momentum scales $\gsim
\Lambda_{TC}$, are defined at that scale. There are a variety of other quantum
corrections to the physical mass and dipole matrices, coming from momentum
scales $\lsim \Lambda_{TC}$ and involving virtual particles with masses in this
range. These arise from standard-model interactions and iterations of the
higher-dimension operators (e.g., \cite{tgv}). We focus here on the
contribution of physics at scales $\gsim \Lambda_{TC}$ incorporated in the
above operators.

The mass operator (\ref{leptonmass}) may be estimated from a graph in which an
incoming lepton $\ell_{k}$ goes to an internal technifermion and ETC vector
boson that has become massive at the stage in the breaking of the full ETC
gauge group to SU(2)$_{TC}$ corresponding to the generation index $k$. The
technifermion develops a soft dynamical mass as the technicolor interactions
break the associated chiral symmetry, and it then recombines with an ETC vector
boson to give the outgoing lepton $\ell_j$. There is in general complex mixing
among the ETC group eigenstates \cite{at94,nt,lrs,ckm} arising at the various
stages of ETC breaking, producing complex off-diagonal terms in the mass matrix
$M^{(\ell)}_{jk}$.

The mass matrix $M^{(\ell)}_{jk}$ may be expressed as
\beq M^{(\ell)}_{jk} \simeq \frac{8 \pi \Lambda_{TC}^3 \Pi_{jk} \ \eta}
{3 \Lambda_j^2 \Lambda_k^2} \label{massoperator} \eeq
where $\Lambda_{TC}$ is the technicolor confinement scale, $\Lambda_{j}$ is the
ETC scale associated with the family index $j$, and $\Pi_{jk}$ is a complex
function of the ETC scales with mass-squared dimensions arising from the ETC
gauge boson mixing.  The $8\pi/3$ factor is derived in Ref.  \cite{ckm}; this
factor and coefficients of other expressions herein are subject to
uncertainties due to the strong-coupling nature of the ETC interactions. The
magnitude of $\Pi_{jk}$ is no greater than min($\Lambda_j^2,\Lambda_k^2$), and
hence may be treated perturbatively through a single insertion. The factor
$\eta$ is $O(1)$ in a QCD-like technicolor theory, while in a theory with
walking from $\Lambda_{TC}$ to the lowest ETC scale $\Lambda_3$ and with
anomalous dimension 1, $\eta = O(\Lambda_3/\Lambda_{TC})$. The off-diagonal
terms in $M^{(\ell)}_{jk}$ determine the structure of $U_L$ and $U_R$, the
former entering along with the corresponding transformation connecting the
neutrino interaction and mass eigenstates in the observed lepton mixing matrix.

To estimate $\tilde D$, we note that the relevant graph is obtained from the
corresponding mass graph by the coupling of a photon to the internal
technifermion \cite{neutrinodipole}. It is more convergent than the mass graph
by two powers of the loop momentum. The fact that the necessary technifermion
mass is soft above $\Lambda_{TC}$ then leads to the convergence of the momentum
integral at momenta of this order -- well below the ETC scales. Thus a
potential contribution of order $1/\Lambda^2_{ETC}$ may be estimated by setting
to zero the momentum flowing through the ETC gauge boson propagator including
the mixing; this effectively replaces the ETC vector boson exchange by a
four-fermion interaction, which does not yield a contribution to the dipole
moment operator.

The leading contribution to $\tilde D_{jk}$ therefore has additional inverse
powers of the ETC mass scales and arises from integration momenta on this
order. The resultant integral producing this leading term has the same power
counting as the integral for $M_{jk}^{(\ell)}$ (\ref{massoperator}) in the
momentum range from $\Lambda_{TC}$ up to the lowest relevant ETC scale, but at
this scale and above it is quite different in detail. Taking account of
multiple ETC scales for different generations, the elements of the
dipole matrix $\tilde D_{jk}$ are
\beq \tilde D_{jk} \simeq \frac{e M_{jk}^{(\ell)}}{\Lambda_{jk}^2}
\label{dmrel} \eeq
where $\Lambda_{jk}$ is of order the scale at which the relevant ETC propagator
including the mixing function becomes soft. It is no greater than ${\rm
min}(\Lambda_j, \Lambda_k)$, and can be less.

We note that Eqs. (\ref{massoperator}) and (\ref{dmrel}) can be obtained also
using an effective ETC theory, employing, for $E < \Lambda_j$, local operators
of dimension-six and higher. Dimension-six (four-fermion) operators generate
$M_{jk}^{(\ell)}$ while dimension-eight operators (four-fermion with derivative
couplings) similarly generate $\tilde D_{jk}$.  However, ETC gauge theories,
such as ours, operative at ETC scales and above, provide more information about
the scales in Eqs. (\ref{massoperator}) and (\ref{dmrel}).

Since $\tilde D_{jk}$ is not, in general, $\propto M_{jk}^{(\ell)}$, it is not
diagonalized by the transformation that diagonalizes $M^{(\ell)}$. The
transformation yields instead the (non-diagonal and complex) dipole matrix $D$
(\ref{ddtilderel}). A principal result of our analysis is that mixing
generically has an important effect on charged-lepton dipole moments in ETC
theories. This is true even for relatively small mixing.

For numerical estimates of the dipole matrix, we take $\Lambda_1 \simeq 10^3$
TeV, $\Lambda_2 \simeq 10^2$ TeV, and $\Lambda_3 \simeq 4$ TeV, as in
Refs. \cite{nt,ckm}. These values can yield realistic ranges for quark and
lepton masses, since for down quarks and charged leptons there can be a natural
suppression so that $|\Pi_{jk}| \lsim {\rm min}( \Lambda_j^2,\Lambda_k^2)$. A
mechanism for this suppression, using relatively conjugate ETC representations
for these fields, is given in \cite{ckm}, although none of the models yet
yields exactly the requisite values for the $\Pi_{jk}$'s.

Each of the matrices $U_\chi$, $\chi=L,R$, depends on three rotation angles
$\theta_{mn}^\chi$, $mn =12,13,23$, and six phases $e^{i\alpha_j^\chi}$,
$e^{i\beta_j^\chi}$, $j=1,2,3$.  We use the conventions of \cite{pdg} for the
$\theta^{\chi}_{mn}$ and write $U_\chi = P_\alpha^\chi
R_{23}(\theta_{23}^\chi)R_{13}(\theta_{13}^\chi)
R(\theta_{12}^\chi)P_\beta^\chi$, where $R_{mn}(\theta_{mn}^\chi)$ is the
rotation through $\theta_{mn}^\chi$ in the $mn$ subspace and $P_\alpha^\chi =
{\rm diag}(e^{i \alpha_1^\chi}, e^{i\alpha_2^\chi}, e^{i\alpha_3^\chi})$.  The
$D_{jk}$ are independent of $P_\beta^\chi$, $\chi=L,R$. The rotation angles
are model-dependent, typically small if the off-diagonal $\Pi_{jk}$'s are more
suppressed than the diagonal ones, and large if this is not the case.

For the electron EDM, using Eqs. (\ref{ddtilderel}) and (\ref{dmrel}), we find
\beq
\frac{d_{e}}{e} \simeq \frac{m_\tau {\rm Im}(F_{11,3})}{\Lambda_3^2}
\label{de} \eeq
where we have kept only the term with the largest ($\tau$) lepton mass in the
numerator and the smallest ETC scale in the denominator. Here, $F_{11,3}$ is a
dimensionless function of the parameters in $U_L$ and $U_R$, of $O(1)$ for
generic values of these parameters, which vanishes if all phases are 0 mod
$\pi$ or if mixing angles vanish. The complex phases remain in $D =
U_{L}\tilde{D}U_{R}$ because of the non-proportionality of $\tilde D$ and
$M^{(\ell)}$. If mixing were absent, the phases in $\tilde D$ would be the same
as the phases in $M^{(\ell)}$ (both diagonal) and would be removed by the
transformation that makes the latter real. In a series expansion in small
rotation angles up to quadratic order, $F_{jk,3} =
e^{i[(\alpha_j^L-\alpha_3^L)-(\alpha_k^R-\alpha_3^R)]}
\theta_{j3}^L\theta_{k3}^R$ for $j,k \ne 3$.

The current upper limit on the electron EDM is $|d_e| < 1.6 \times 10^{-27}$
e-cm \cite{regan}, and ongoing experiments project sensitivities down to
$10^{-30}$ e-cm or better \cite{eedm_exp}. Comparing Eq. (\ref{de}) to the
upper limit, with $\Lambda_3 \simeq 4$ TeV, we see that Im($F_{11,3}$) must be
much less than $O(1)$; in fact, Im($F_{11,3})$ $\lsim 0.7 \times
10^{-6}$. Taking the phases to be generic, of $O(1)$, we conclude that the
mixing angles must be small. To bound them, we use the above expression for
$F_{jk,3}$, neglecting terms beyond quadratic order in the products of the
various $\theta$'s. (Higher-order terms can be included in a more detailed
analysis.)  We then have $|\theta_{13}^L\theta_{13}^R| \lsim 10^{-6}$
\cite{ae}. Values of the angles in this range are not unexpected, given the
suppression of the off-diagonal mixings $\Pi_{jk}$ in some of the models of
Refs. \cite{nt,ckm}. Even if the product $|\theta_{13}^L\theta_{13}^R|$ is
somewhat below the upper bound, the ETC contribution to $d_e$ can naturally lie
in the range accessible to ongoing experiments.

The off-diagonal elements $D_{jk}$ produce flavor-changing radiative lepton
decays. From the upper bounds on these, in conjunction with the above values
of the ETC scales $\Lambda_j$ chosen to yield appropriate fermion masses, we
can derive additional bounds on charged lepton mixing. We have
$\Gamma(\psi_k \to \psi_j \gamma) =
(|D_{jk}|^2+|D_{kj}|^2)m_{\psi_k}^3/(8\pi)$.  For example, for the case
$k=2$, $j=1$, i.e., $\mu \to e \gamma$, the terms with the dominant ETC scale
dependence are
\beq  \frac{D_{jk}}{e} \simeq \frac{m_\tau F_{jk,3}}{\Lambda_3^2} \ , \quad jk
= 12, \  21 \ .
\label{djkterms} \eeq
We infer an upper bound from the limit $B(\mu \to e \gamma) < 2.1 \times
10^{-11}$ \cite{pdg,psi}, viz., $|\theta_{13}^L \theta_{23}^R|, \
|\theta_{13}^R \theta_{23}^L| \lsim 5 \times 10^{-6}$. Again, this range of
values is consistent with some models in \cite{nt,ckm,muecon}.

Since $U_L$ enters along with neutrino mixing into the observed lepton
mixing in neutrino oscillations, our bounds on $\theta^{L}_{jk}$ suggest
\cite{msym}, that mixing in the neutrino sector is the primary source of
the large measured lepton mixing angles $\theta_{23}$ and $\theta_{12}$.
Large neutrino-sector mixing can emerge naturally from some of the models of
Refs. \cite{nt,ckm}. It is also the case that large leptonic CP violation,
as could be observed in future neutrino oscillation experiments, is natural
in these models.

 From the limits $B(\tau \to \mu \gamma) < 3.1 \times 10^{-7}$ \cite{tmg} and
$B(\tau \to e \gamma) < 2.6 \times 10^{-7}$ \cite{pdg} we obtain the
respective bounds $|\theta_{23}^L|, \ |\theta_{23}^R| \lsim 0.02$ and
$|\theta_{13}^L|, \ |\theta_{13}^R| \lsim 0.06$. The linear form of these
bounds is due to the fact that one of the external particles is the $\tau$.
The dominant terms in the amplitude for the respective decays, in a
small-$\theta$ expansion, are $D_{j3} \simeq e^{i(\alpha_j^L-\alpha_3^L)}
\theta_{j3}^Lm_\tau/\Lambda_3^2$ and $D_{3j} \simeq
e^{i(\alpha_3^R-\alpha_j^R)}\theta_{j3}^R m_\tau/ \Lambda_3^2$, with
$j=2,1$.

For the muon $g-2$, keeping the dominant terms,
\beq \frac{a_\mu}{2 m_\mu} \simeq \frac{m_\mu}{\Lambda_{22}^2} + \frac{
m_\tau {\rm Re}(F_{22,3})}{\Lambda_3^2} \ , \label{amu} \eeq
where $\Lambda_{22}$ is the softness scale of the relevant ETC exchange,
ranging, in the models explored, from $\Lambda_3$ ($\simeq 4$ TeV) to
$\Lambda_2$ ($\simeq 10^2$ TeV). The first term in (\ref{amu}) would be present
even without mixing \cite{cm}. For $\Lambda_{22} \simeq \Lambda_2$ and
$|\theta_{23}^L|$ and $|\theta_{23}^R|$ bounded as above, the second term can
dominate the first, but the ETC contribution to $a_\mu$ is $\lsim 10^{-11}$,
well below the current uncertainty $\sim 10^{-9}$ in the comparison of theory
and experiment \cite{e821}. However, if $\Lambda_{22} = O(\Lambda_{3})$ as in
some models, the first term would dominate and, interestingly, would be of
order $10^{-9}$.

For the muon EDM, we find
\beq
\frac{d_{\mu}}{e} \simeq \frac{m_\tau {\rm Im}(F_{22,3})}{\Lambda_3^2} \ .
\label{dmu}
\eeq
With $|\theta_{23}^L\theta_{23}^R|$ near the upper limit $4 \times 10^{-4}$, we
estimate that $|d_{\mu}|$ could be $\simeq 10^{-24}$ e-cm. This is well below
the current limit $|d_\mu| < 3.7 \times 10^{-19}$ e-cm \cite{pdg,feng} but
might be observable in the proposed experiment of \cite{semertzidis}.

For the $\tau$-lepton, the bounds $|a_\tau| < 0.06$ and $|d_\tau| < 3.1 \times
10^{-16}$ e-cm \cite{pdg} are not sensitive to the contributions described
here, arising from physics at scales $\ge \Lambda_{TC}$.

In summary, we have analyzed the magnetic and electric dipole moments of
charged leptons in a class of ETC models with lepton mixing and dynamically
generated  CP-violating phases. We have shown that the ETC contribution to
the electron EDM is dominated by terms from charged lepton mixing and can be
comparable to the current experimental limit. We have used current limits on
$|d_e|$ and radiative lepton decays to set bounds on charged lepton mixing
angles. We have noted that these constraints are such that the ETC
contribution to the muon anomalous magnetic moment, which includes a
significant lepton mixing term, can approach, but does not exceed, the
current sensitivity level.

We thank Kenneth Lane and David DeMille for helpful comments.  This research
was partially supported by the grants DE-FG02-92ER-4074 (T.A., M.P.) and
NSF-PHY-00-98527 (R.S.).


\begin{thebibliography}{99}
\bibitem{msym}
In the models of Refs. \cite{nt,lrs,ckm} $|M_{jk}^{(\ell)}| =
|M_{kj}^{(\ell)}|$ (hermitian or complex symmetric) so that $U_L$ and $U_R$
are related. We keep our present discussion more general.
\bibitem{neutrinodipole} Since the ETC gauge bosons are neutral, there is no
contribution to neutrino dipole moment matrices. Also, since in \cite{nt,ckm}
the neutrinos are generically Majorana particles, the diagonal dipole moments
($\mu_\nu$ and $d_\nu$) vanish.
\bibitem{ae}
These mixings also enter $a_{e,ETC}$ so that it is far below the
experimental accuracy $\Delta a_e = 4 \times 10^{-12}$ \cite{pdg}.
\bibitem{muecon} We also considered limits from bounds on $\mu \to e$
conversion; these are less stringent than those from $\mu \to e \gamma$.
\end{thebibliography}
\end{document}